\documentclass[aip,jcp,reprint,showkeys,superscriptaddress,longbibliography,noeprint]{revtex4-1}
\usepackage[utf8]{inputenc}
\usepackage[english]{babel}
\usepackage[utf8]{inputenc}
\usepackage{systeme}
\usepackage[fleqn]{amsmath}
\usepackage{amssymb}
\usepackage{mleftright}
\usepackage{siunitx}
\usepackage[inline,shortlabels]{enumitem}

\usepackage{soul} % highlight text: \hl command

\newcommand{\recheck}[1]{\bgroup\sethlcolor{yellow}\hl{#1}\egroup}

\usepackage{graphicx,grffile,subfigure}
\graphicspath{{../figures/}}
\newlength\figwidth
\usepackage{url,hyperref}
\usepackage[table,usenames,dvipsnames]{xcolor}
\hypersetup{colorlinks=true, linkcolor=BrickRed,
  urlcolor=blue!50!black, citecolor=blue!50!black}
\usepackage[capitalize]{cleveref}
\crefrangelabelformat{equation}{(#3#1#4) to (#5#2#6)}

\usepackage{booktabs,tabularx,dcolumn}
\newcolumntype{d}[1]{D{.}{.}{#1}}

\makeatletter
\def\paragraph{%
  \@startsection
    {paragraph}{4}{\parindent}{\z@}{-1.5em}%
    {\normalfont\normalsize\itshape}%
}%
\makeatother

\renewcommand\epsilon{\varepsilon}
\renewcommand\phi{\varphi}
\renewcommand\theta{\vartheta}
\renewcommand\rho{\varrho}
\renewcommand\leq{\leqslant}

\renewcommand\vec[1]{\textrm{\bfseries #1}}
\newcommand\expect[1]{\left\langle{#1}\right\rangle}
\newcommand\e{\text{e}}
\newcommand\kB{k_{\text{B}}}
\newcommand\Nbar{{\overline{N}}}
\newcommand\Ad{{\text{Ad}}}
\newcommand\AT{{\text{AT}}}

\newcommand\at{{\text{at}}}
\newcommand\res{r} %{{\text{res}}}
\renewcommand\th{{\text{th}}}
\newcommand\id{{\text{id}}}
\newcommand\ex{{\text{ex}}}
\DeclareMathOperator\std{std}

\makeatletter\AtBeginDocument{\let\@elt\relax}\makeatother

\begin{document}

%\blue{\noindent\footnotesize\itshape Comments in blue are turned off by uncommenting line 18 of the LaTeX source.}

\title{Thermodynamic relations at the coupling boundary in adaptive resolution simulations for open systems}
% \title{Thermodynamic relations at the coupling boundary in the open system/adaptive resolution technique}

\newcommand\FUBaffiliation{\affiliation{Freie Universität Berlin, Institute of Mathematics, Arnimallee 6, 14195 Berlin, Germany}}
\author{Abbas Gholami}
\FUBaffiliation
\author{Felix H\"{o}f{}ling}
\FUBaffiliation
\affiliation{Zuse Institute Berlin, Takustr. 7, 14195 Berlin, Germany}
\author{Rupert Klein}
\FUBaffiliation
\author{Luigi Delle Site}
\email{luigi.dellesite@fu-berlin.de}
\FUBaffiliation

\begin{abstract}
The adaptive resolution simulation (AdResS) technique couples regions with different molecular resolutions and allows the exchange of molecules between different regions in an adaptive fashion. The latest development of the technique allows to abruptly couple the atomistically resolved region with a region of non-interacting point-like particles. The abrupt set-up was derived having in mind the idea of the atomistically resolved region as an open system embedded in a large reservoir at a given macroscopic state. In this work, starting from the idea of open system, we derive thermodynamic relations for AdResS which justify conceptually and numerically the claim of AdResS as a technique for simulating open systems. In particular, we derive the relation between the chemical potential of the AdResS set-up and that of its reference fully atomistic simulation. The implication of this result is that the grand potential of AdResS can be explicitly written and thus, from a statistical mechanics point of view, the atomistically resolved region of AdResS can be identified with a well defined open system.
\end{abstract}

\maketitle

% ===============================================================================================
% ===============================================================================================
% ===============================================================================================

\section{Introduction}

The adaptive resolution simulation (AdResS) technique couples, in a concurrent fashion, regions of space at different molecular resolutions \cite{jcpor,annurev,physrep}. Recent developments are pushing the method towards a computational realization of an open system embedded in a large reservoir of particles and energy \cite{leb1,leb2,njp,preliouv,krekjcp, advtscomm,jmp,noneq1}. The simulation set-up is reduced to the very essential by abruptly coupling an atomistically resolved region to a reservoir of non-interacting point particles \cite{advtscomm}. The simplified algorithmic protocol, explained in detail in the next section, has the advantage of high computational efficiency and allows us to writing a total interaction potential without making use of artificial, space-dependent, interpolations of atomistic and coarse-grained forces or Hamiltonians \cite{jcpor,prlraff,prlmatej}. The abrupt coupling between the different regions may give the impression, at a first glance, of being highly artificial; in reality, physical consistency can be achieved by imposing specific numerical conditions. These latter assure that the AdResS simulation reproduces the results of the simulation of an equivalent subsystem in a large fully atomistic system of reference \cite{jctchan, prx}.

In this perspective, the natural question arising is whether one can translate the numerical constraints in explicit thermodynamic and statistical mechanics relations occurring at the coupling region. This work demonstrates that arguments relying on physical consistency indeed lead to explicit thermodynamic descriptions of the AdResS set-up that positively pass specific numerical tests. The key result of the paper is the relation between the chemical potential of the atomistically resolved subsystem and the chemical potential of the fully atomistic system of reference. Such a relation, in turn, allows one to define the grand potential of the atomistic region of AdResS in terms of quantities that can be explicitly calculated from numerical simulations. The grand potential expresses the essential thermodynamic and statistical mechanics features of an open system. Thus the possibility of concretely defining the grand potential of AdResS at the microscopic level provides a robust justification to the idea of AdResS as a physically consistent numerical approach to open systems. The derivation of the thermodynamic relation is developed under ideal conditions which do not normally occur in standard simulations; however, numerical tests suggest that the obtained relations can be applied beyond the ideal conditions in which they have been derived.
The results of this paper enrich the thermodynamic and statistical mechanics foundations of AdResS in its abrupt coupling approach and stimulate future deeper analysis of its several theoretical and numerical implications.
% The paper is organized as follows: next section describes the basic features of the AdResS protocol in its latest version of abrupt coupling. Next, the derivation of the thermodynamic relation between the chemical potential of the AdResS system and the chemical potential of the reference fully resolved system is reported. It follows the section with the numerical results and a final discussion concludes the paper.

\section{Basic principles of $\text{AdResS}$ with a reservoir of non-interacting particles}

\begin{figure}
  \centering\includegraphics[clip=true,trim=0.1cm 0cm 0cm 0.1cm,width=\figwidth]{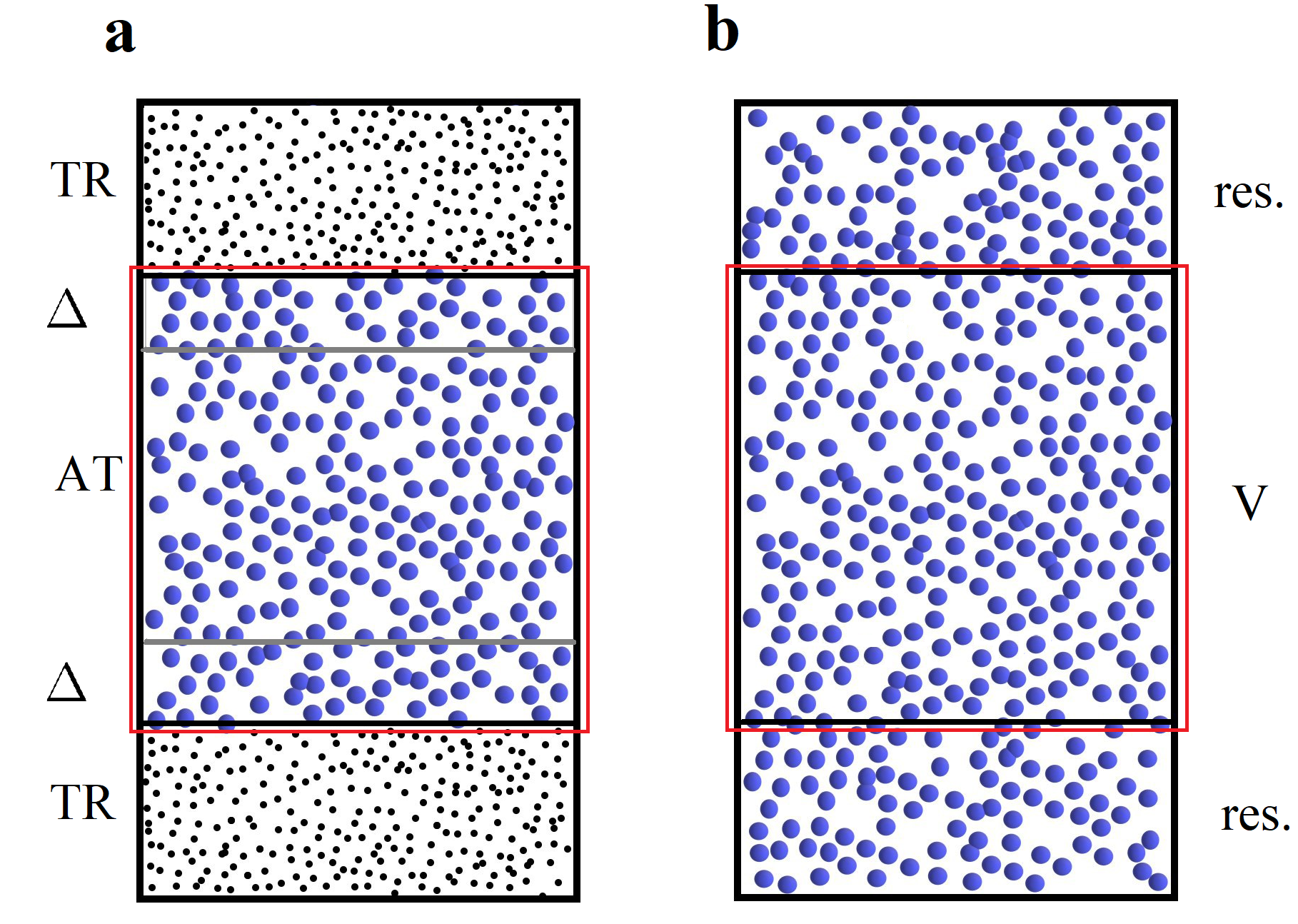}
  \caption{Simulation set up in the (a) AdResS system and (b) fully atomistic system. In AdResS particles change resolution when crossing the border between the $\Delta$ and TR region (reservoir). In both systems, the red boxes represent the subsystem {analyzed} in this work. { It must be underlined that the AT region is the region of physical interest, the $\Delta$ region is an AdResS-artifact through which the coupling to the non-interacting particle reservoir becomes technically possible. Here we extend the analysis to the $\Delta$ region so that its coupling conditions can be rationalized in terms of thermodynamic quantities of the joint $\AT\cup \Delta$ region.}}
  \label{cartoon}   
\end{figure}

\Cref{cartoon} illustrates the AdResS set-up; this latter consists of partitioning the simulation box into three regions: the region of interest AT, with fully atomistic resolution, the interface region $\Delta$, with fully atomistic resolution but with additional coupling features to the large reservoir, and TR, the large reservoir of non-interacting particles.
Molecules of the AT region interact with atomistic potentials among themselves and with molecules in $\Delta$, and vice versa, while there is no direct interaction with the tracer particles.
Tracers and molecules in the $\Delta$ and TR regions are subject to an additional one-body force $F_\th(x)\, \vec n$, named thermodynamic force, acting along the direction $\vec n$ in which the change of resolution takes place; it is as function of the distance $x$ from the atomistic region and $\vec n$ is the surface normal of the coupling boundary.
Second, a thermostat acts on the $\Delta$ and TR regions that compensates the heat introduced by the change of resolution \cite{prx,noneq1}.
In essence, these are the coupling condition between the $\Delta$ region and the reservoir TR.

Technically, also a force capping is imposed in the $\Delta$ region since point-like particles arriving from the TR region and entering the $\Delta$ region may be unphysically close to one other. Due to the abrupt switching of molecular degrees of freedom, close molecules can experience forces between atoms which are artificially large. Admittedly the {force capping} is an artificial means by which unphysically large forces are automatically relaxed to the average force occurring in the equivalent fully atomistic simulation.
The capping, however, is equivalent to a global modification of the highly repulsive part of the interaction potentials, which has marginal repercussions on the physical properties of the fluid.
The exact form of the force capping is given in the Appendix, where we also report numerical tests showing that its effects can be neglected.

In summary, the total potential of the AdResS set-up reads
\begin{equation}
U_\Ad(x_N) = U(x_N) + \Phi_\Delta(x_N) + U_\mathrm{cap}(x_N)
\label{total_potential}
\end{equation}
assuming that at a given instance in time $N$ particles are found in the $\AT \cup \Delta$ region with positions $x_N=\{\vec r_1, \dots, \vec r_N\}$.
Here, $U(x_N)$ represents the total potential from atomistic interactions of particles in $\AT\cup \Delta$ among themselves;
$\Phi_\Delta(x_N) := \sum_{\vec r_j\in \Delta} \phi_\th(\vec r_j)$
collects the contributions due to the potential $\phi_\th$ of the thermodynamic force, $F_\th(\vec r) = -\nabla \phi_\th(\vec r)$ with $\phi_\th=0$ at the AT/$\Delta$ interface \cite{advtscomm}. Finally, $U_\mathrm{cap}(x_N)$ arises from the force capping and is only present in the $\Delta$ region.

The effect of $\vec{F}_\th(r)$ consists in enforcing a homogeneous molecular density in the $\Delta$ region equivalent to the molecular density $\rho_\at$ in equilibrium of the reference fully atomistic system.
In practice, it is calculated self-consistently in an iterative process, starting from
$F_\th^{(0)}(x) = 0$. The update between sucsessive steps $k$ is
$F_\th^{(k+1)}(x) = F_\th^{(k)}(x) - c\, \nabla \rho_k(x)$, % (M/\rho_0^2 \kappa_T)
where the density profile $\rho_k(x)$ was calculated from an AdResS simulation using $F_\th^{(k)}(x)$
and $c > 0$ is a suitable coefficient to control the speed of convergence.
The iteration stops when the deviation of $\rho_k(x)$ from a constant profile is within a prescribed tolerance, details are given in the Appendix.
After $F_\th(x)$ has been determined, it remains unchanged in the whole AdResS production run without recalibration \cite{prl2012,prx}.

The development of the abrupt computational set-up with tracers became possible through the mapping of the algorithm onto a theoretical model of open systems. Such a model fixes a series of conditions that the AdResS simulation must fulfill to be considered valid. Such conditions are sufficient to assure that the physics of the AT region is correct and, in the limit of large TR and AT regions compared to $\Delta$, one has a Grand Canonical-like ensemble for AT (GC-AdResS)\cite{softmatt}. The application of the thermodynamic force in $\Delta$ is one of these conditions because it assures that the particle density in $\Delta$ is equal to the atomistic target density at equilibrium. {Ideally one would want to make sure that the interactions of particles near the boundary of the AT region with their neighbors are statistically isotropic, i.e. independent of whether neighbors are located within the AT or $\Delta$ region. Matching the densities between the AT and $\Delta$ region is a necessary condition for achieving this.}

Going one step further, it is required that the probability distribution $p(N)$ of the number of particles in the AT region should be the same, within some accuracy, as $p(N)$ of the equivalent subsystem in a fully atomistic simulation. The fulfillment of such condition assures one that, in average, the exchange of particles between the AT region and the reservoir occurs in the proper manner.
For a Gaussian distribution, it is sufficient to compare the first two cumulants, which are related to the density and the compressibility, where the latter provides a particular sensitive test of the boundary conditions \cite{jcp-fscorr}.
In the language of statistical mechanics, the equivalence of the $n$-th moments of $p(N)$ between AdResS and the reference system guarantees that the $n$-th derivatives of the grand potentials of the two systems with respect to the chemical potential agree and vice versa.
Additionally, microscopic structural consistency is assured by matching atom--atom radial distribution functions in the AT region.
% which must agree with the equivalent functions calculated in the same subsystem of a fully atomistic simulation of reference.
Finally, one could also verify that the interaction energy between molecules in AT and molecules in $\Delta$ is negligible compared to the interaction energy amongst the molecules in AT, so that the physics of the system is determined only by the interactions between the molecules in AT, that is the physical system of interest.

\section{Relation between the chemical potential of $\text{AdResS}$ and of a fully atomistic system of reference}

\subsection{Principle of equivalence for the grand potential}

In an open system, the relevant thermodynamic state potential is the grand potential
$\Omega = -PV$, where $V$ and $P$ denote the volume and the pressure of the system.
It has the microscopic expression
\begin{equation}
\Omega = -\kB T \ln \left(\sum_{N=0}^\infty \e^{\beta \mu N} {Q_N}\right),
\label{gp}
\end{equation}
if the system is equilibrated at the chemical potential $\mu$ and the temperature $T$;
as usual $\beta=1/k_{B}T$ with $k_{B}$ Boltzmann's constant.
The partition function at fixed number $N$ of identical particles reads
\begin{equation}
Q_N = \frac{1}{h^{3N}N!} \int_{\mathbb{R}^{3N}}\int_{V^{N}}e^{-\beta H_N(x_N, p_N)}dx_{N}dp_{N},
\label{qn}
\end{equation}
where $p_{N}=\{\vec p_{1},\dots,\vec p_{N}\}$ and $x_{N}=\{\vec r_{1},\dots, \vec r_{N}\}$ are the momenta and positions of the $N$ particles, respectively. The Hamiltonian is the sum of kinetic and interaction potential energies:
\begin{equation}
H_N(x_N, p_N) = \sum_{i=1}^N \frac{\vec{p}_i^{2}}{2m} + U(x_N),
\label{hamiltonian}
\end{equation}
and $m$ is the particle mass.
It must be noticed that here the potential $U(x_N)$ contains
% ideally considers
interactions only between particles in the system and neglects any potential interaction with the exterior (see e.g. Ref. \citenum{huang}).

For a subsystem $S$ in a fully atomistic system of reference, whose domain is equivalent to the $S=\AT \cup \Delta$ region of the AdResS set-up, let us define the grand potential of the reference system as
\begin{equation}
 \langle\Omega_{r}\rangle = -\kB T \ln \left(\sum_{N=0}^\infty \e^{\beta \mu_{r} N} {Q^{r}_N} \right)\,,
 \label{gp-ref}
\end{equation}
where we denoted the chemical potential of the reference subsystem by $\mu_r$ and introduced the effective $N$-particle partition function
\begin{equation}
{Q^{r}_N}:=\left\langle \frac{1}{h^{3N}N!} \int_{\mathbb{R}^{3N}}\int_{S^{N}}e^{-\beta H^{r}_N(x_N, p_N|x'_M)} dx_{N}dp_{N}\right\rangle_{\delta S} .
  \label{orm}
\end{equation}
Here, the extended Hamiltonian
\begin{equation}
  H^{r}_{N}(x_N, p_N | x'_M) = \sum_{i=1}^N \frac{\vec{p}_i^{2}}{2m} + U(x_N)+U(x_{N},x'_{M})
\end{equation}
expresses the fact that the $N$ molecules of the subsystem do not interact only among themselves, but also with $M$ molecules located in a layer $\delta S$ around the $S$ region.
The angular brackets in \cref{orm} denote an averaging operation over the positions $x'_M \in \delta S$ of these reservoir particles,
which, however, are correlated with other particles of the reservoir outside of $\delta S$.
Mathematically, the probability density $p_{\delta S}(x'_M)$ of the positions in $\delta S$ is obtained by marginalization of the phase space density of the universe (subsystem plus reservoir), see also Ref.~\citenum{jmp}.
% The averaging operation in \cref{orm} is equivalent to a marginalization w.r.t. the outer particles which essentially gives the statistical average of $\Omega_{r}$ over all possible $x'_{M}$ states (see also \cite{jmp}).
Later on we will actually specify how one performes the marginalization w.r.t. the outer particles in a simulation where $M$ changes dynamically.
If we assume that $U(x)$ is a potential with a sufficiently short interaction range, so that the volumes obey $|\delta S|\ll |S|$, and thus the integration over the $M$ particles represents a surface effect, we can, in good approximation, identify $\langle\Omega_{r}\rangle$ with the grand potential of the $S$ region (see also Ref.~\citenum{huang}).

Next, we consider the $S=\AT \cup \Delta$ region of the AdResS set-up \emph{without} the thermodynamic force acting in the $\Delta$ region, i.e., without the potential energy contribution $\phi_\th(x)$.
Denoting by $\mu_{0}$ the chemical potential of this subsystem in absence of the thermodynamic force,
and the corresponding grand potential $\Omega^{0}_\Ad$ is defined as
\begin{equation}
  \Omega^{0}_\Ad=-\kB T \ln \left(\sum_{N=0}^\infty \e^{\beta \mu_{0} N} {Q_N^{\Ad,0}} \right)
\end{equation}
with
\begin{equation}
{Q_N^{\Ad,0}}=\frac{1}{h^{3N}N!} \int_{\mathbb{R}^{3N}}\int_{S^{N}}e^{-\beta H_{N}^{\Ad,0}(x_N, p_N)}dx_{N}dp_{N}
\end{equation}
and the Hamiltonian
\begin{equation}
H_{N}^{\Ad,0}(x_N, p_N)=\sum_{i=1}^N \frac{\vec{p}_i^{2}}{2m} + U(x_N) + U_\mathrm{cap}(x_{N}) \,;
\end{equation}
the latter follows from \cref{total_potential}. As argued above, the number of capping events is negligible and we will neglect the term $U_\mathrm{cap}(x_{N})$ in the following, so that $H_{N}^{\Ad,0}$ reduces to $H_{N}$ as in \cref{hamiltonian}.

The purpose of AdResS is to reproduce the physics of the reference fully atomistic simulation in the AT region. If the AdResS set-up, with Hamiltonian $H_{N}^{\Ad,0}$, was sufficient to this aim one should have: $\langle\Omega_{r}\rangle=\Omega^{0}_\Ad$, but it is easy to numerically verify that this is never the case. However, as described in the previous section, adding a one-particle potential in the $\Delta$ region of the AdResS set-up is sufficient to enforce the physical consistency between AdResS and its reference fully atomistic system.
In the following subsection, we will interpret the inclusion of the potential of the thermodynamic force in AdResS through the idea of equivalence of the grand potential between AdResS and its fully atomistic system of reference.

\subsection{Perturbation of the potential energy in the $\Delta$ region}

Let us anticipate the thermodynamic limit so that the size of AT is arbitrarily large and $|\AT|\gg |\Delta|$. Under such conditions, we add a small perturbation to the potential of the $\Delta$ region in AdResS. Let us assume that such a perturbation can be designed in such a way that we achieve the wished relation of thermodynamic equivalence between the AdResS set-up and its reference simulation:
\begin{equation}
\langle\Omega_{r}\rangle=\Omega_\Ad \,;
\label{gp-equiv}
\end{equation}
equality holds also for all derivatives of the two grand potentials w.r.t.\ the variables $\mu, V, T$.

In the actual AdResS numerical simulation, the potential of the thermodynamic force $\phi_\th(x)$ acting in $\Delta$ assures the approximate statistical equivalence between the AdResS simulation and its fully atomistic simulation of reference within AT, at least for the one-particle density and the pair (radial) distribution function. Thus, $\phi_\th(x)$ is a reasonable approximation to the perturbation needed. In the presence of a perturbation, one can assume that physical quantities of interest in $\AT\cup\Delta$ remain, in first approximation, as they were in absence of the perturbation and that the effect of the perturbation can be explicitly derived and added to them. This argument allows us to write in good approximation for the grand potential of the equilibrated AdResS set-up:
\begin{equation}
\Omega_\Ad=-\kB T \ln \left(\sum_{N=0}^\infty \e^{\beta (\mu_{0}+\Delta\mu) N} {Q_{N}^{\Ad}} \right)
\label{gp-adress}
\end{equation}
with
\begin{equation}
{Q_{N}^{\Ad}}=\frac{1}{h^{3N}N!} \int_{\mathbb{R}^{3N}}\int_{S^{N}}e^{-\beta H_{N}^{\Ad}(x_N, p_N)}dx_{N}dp_{N} \,,
\label{qnad}
\end{equation}
in which $H_{N}^{\Ad}(x_N, p_N) = H_{N}^{\Ad,0}(x_N, p_N) + \Phi_\Delta(x_{N})$.
% \begin{equation}
% H_{N}^{\Ad}(x_N, p_N)=\sum_{i=1}^N \frac{\vec{p}_i^{2}}{2m} + U(x_{N})+\Phi_\Delta(x_{N}) \,.
% \end{equation}
Here, we have denoted the chemical potential of the perturbed system by $\mu_{0}+\Delta\mu$ and assumed, according to the above definition of perturbation, that the difference $\Delta\mu=\Delta\mu[\phi_\th]$ originates from the perturbation of the potential energy in $Q_{N}^{\Ad}$.
In order to arrive at an explicit thermodynamic relation between $\mu_{r}$, $\mu_{0}$ and $\Delta\mu$, we will derive explicit expressions of $Q_{N}^{r}$ and $Q_{N}^{\Ad}$ in the subsection below.

\subsection{Relation between $\mu_{r}$, $\mu_{0}$ and $\Delta\mu$}

Following the standard arguments in statistical mechanics\cite{huang}, the sum over $N$ in the expressions \cref{gp-ref,gp-adress} for $\langle\Omega_{r}\rangle$ and $\Omega_\Ad$ represents a major obstacle to the derivation of a direct relation between $\mu_{r}$, $\mu_{0}$ and $\Delta\mu$. However, given the conditions of the thermodynamic limit for $S$ we can assume that $p(N)$ is sharply peaked around $\bar{N}$, with $\bar{N}$ being the average number of particles in $S$. Under such assumption, the sum over $N$ can be approximated by its most relevant term, that is the term of the series corresponding to $\bar{N}$. It follows that the condition $\langle\Omega_{r}\rangle=\Omega_\Ad$ [\cref{gp-equiv}] implies:
\begin{equation}
  -\kB T\ln\left(e^{\beta\mu_{r}{\bar{N}}}Q_{\bar{N}}^{r}\right)=-\kB T\ln\left(e^{\beta(\mu_{0}+\Delta\mu){\bar{N}}}Q_{\bar{N}}^{\Ad}\right) ,
      \label{firstsimp}
    \end{equation}
or equivalently,
 \begin{equation}
  \beta\mu_{r}{\bar{N}}+\ln Q_{\bar {N}}^{r}=\beta\mu_{0}{\bar{N}}+\beta\Delta\mu {\bar{N}}+\ln Q_{\bar{N}}^{\Ad} \,,
      \label{secondsimp}
  \end{equation}
which becomes exact in the thermodynamic limit. The error of the approximation can be estimated by considering $\Nbar \pm \Delta N$ as upper and lower bounds on $\Nbar$ in our calculations, with the standard deviation $\Delta N = \sqrt{\langle N^2 \rangle - \langle N \rangle ^2}$.

In the next step, we will rewrite the expressions for the $\bar N$-particle partition functions $Q_{\bar {N}}^{r}$ and $Q_{\bar{N}}^{\Ad}$.
Let us first consider the partition sum of the equilibrated AdResS setup, given in \cref{qnad}:
\begin{multline}
  Q_{\bar{N}}^{\Ad}=\frac{1}{h^{3{\bar {N}}}{\bar {N}}!} \int_{\mathbb{R}^{3{\bar {N}}}}\int_{S^{{\bar {N}}}}
  \e^{-\beta H_{\bar N}(x_{\bar N}, p_{\bar N})} \\
  \times \e^{-\beta \Phi_\Delta(x_{\bar {N}})}dp_{{\bar {N}}}dx_{{\bar {N}}} \,,
  \label{q2}
\end{multline}
which is nothing else than a quantity proportional to the canonical average of $\e^{-\beta \Phi_\Delta(x_{{\bar {N}}})}$ w.r.t.\ the region $S$, i.e.,
\begin{equation}
Q_{\bar{N}}^{\Ad}= Q_{\bar N} \langle \e^{-\beta \Phi_\Delta(x_{{\bar {N}}})}\rangle\,.
\label{q_adress}
\end{equation}

The evaluation of $Q_{\bar {N}}^{r}$,  defined in \cref{orm}, implies the knowledge of the statistics $p_{\delta S}(x'_M;M)$ of the values of $M$ and the positions $x'_M$ in the shell $\delta S$. For the moment let us assume that it is known and the average w.r.t.\ $x'_{M}$ can be carried out.
For $Q_{\bar N}^r$, this average is spelled out in integral form as
\begin{multline}
  Q_{\bar N}^{r}= \sum_{M=0}^\infty \int_{\delta S^{M}} \frac{1}{h^{3\bar N}\bar N!} \int_{\mathbb{R}^{3\bar N}}\int_{S^{\bar N}}
  \e^{-\beta H_{\bar N}(x_{\bar N}, p_{\bar N})} \\
    \times \e^{-\beta U(x_{\bar N},x'_{M})}
  dx_{\bar N}dp_{\bar N} \,p_{\delta S}(x'_M;M) \,dx'_{M} \,.
  \label{q122}
\end{multline}
This expression can also be interpreted as a canonical average of $\langle \exp(-\beta U(x_{{\bar {N}}},x'_{M})) \rangle_{\delta S}$ over the positions and momenta of the $\bar N$ particles in $S$, after dividing by a suitable normalization factor that coincides with the $\bar N$-particle partition function $Q_{\bar N}$ given in \cref{qn}.
We obtain
\begin{equation}
  Q_{\bar N}^{r} = Q_{\bar N} \left\langle \langle \e^{-\beta U(x_{{N}},x'_{M})} \rangle_{\delta S} \right\rangle
  \label{q_at}
\end{equation}
with the double brackets referring, first, to the average over the statistics of $M$ and the positions $x'_M \in \delta S$ and, second, to the canonical average over $x_{\bar N} \in S$.
In numerical simulations the statistics over $M$ is extracted from a sufficiently long simulation run and the corresponding average over the time series takes properly into account the integration over $x'_{M}$ in \cref{q122}.

Collecting the results of this section and by substituting \cref{q_at,q_adress} in \cref{firstsimp}, one obtains    \begin{equation}
  \e^{\beta \mu_{r} \bar N} \langle \e^{-\beta U(x_{{\bar N}},x'_{M})}\rangle
  = \e^{\beta (\mu_{0} + \Delta\mu)\bar N}  \langle \e^{-\beta \Phi_\Delta(x_{{\bar N}})}\rangle
\end{equation}
and, one step further, an explicit formula that links $\mu_{r}$, $\mu_{0}$ and $\Delta\mu$:
\begin{equation}
  \mu_{r}-(\mu_{0}+\Delta\mu) = \omega_\Delta - \omega_\res \,.
  \label{finaleq}
\end{equation}
Here, the energies
\begin{equation}
  \omega_\Delta := (\beta\bar N)^{-1}\ln{\langle \e^{-\beta \Phi_\Delta(x_{\bar N})}\rangle}
\end{equation}
and
\begin{equation}
  \omega_\res := (\beta\bar N)^{-1}\ln{\langle \e^{-\beta U(x_{\bar N},x'_M)}\rangle},
\end{equation}
are, respectively, the contribution of the potential of thermodynamic force and the pulled out interactions of particles in the open system with those in the reservoir.

Interestingly, $\mu_{r}$, $\mu_{0}$, $\omega_\Delta$, and $\omega_\res$ can be calculated numerically within fully atomistic and AdResS simulations. In particular, $\omega_\Delta$ and $\omega_\res$ contain the terms $\langle \e^{-\beta \Phi_\Delta(x_{\bar N})}\rangle$ and $\langle \e^{-\beta U(x_{\bar N},x'_M)}\rangle$, which, as for the sampling w.r.t.\ the $x'_M$ states discussed above, are calculated by sampling $x_N$ over a sufficiently long trajectory and averaging over the time series. Since we have assumed to work under the condition that $p(N)$ is sharply distributed around $\bar N$, the dominant configurations in the sampling along the trajectory are those with $\bar N$ particles in the subsystem. This means that we can identify with good approximation, $\langle \e^{-\beta \Phi_\Delta(x_{\bar N})}\rangle$ and $\langle \e^{-\beta U(x_{\bar N},x'_M)}\rangle$ with $\langle \e^{-\beta \Phi_\Delta(x_N)}\rangle$ and $\langle \e^{-\beta U(x_N,x'_M)}\rangle$, respectively, calculated from the simulation.

The possibility of calculating numerically the quantities above implies that indeed the grand potential of the AT region of AdResS, $\Omega_\Ad$, within the assumptions made, can be explicitly written. In turn, the explicit definition of the grand potential from a microscopic (first principle of statistical mechanics) perspective legitimates the definition of AdResS as a method of open systems that is well-founded on statistical mechanics.

\section{Numerical experiments}

Numerical experiments to test \cref{finaleq} are performed by molecular dynamics simulations of Lennard-Jones (LJ) fluids for a range of densities so that we gather information for different thermodynamic state points. An additional simulation of liquid water has been carried on to check the applicability of \cref{finaleq} for a system with chemically structured molecules, where the passage from the tracer region to the $\Delta$ region implies the drastic reintroduction of molecular (atomistic) degrees of freedom. Moreover, liquid water is one of the most relevant examples in molecular simulation and AdResS has been shown to handle such systems in a very satisfactory way (see e.g. \cite{advtscomm}) thus it is an ideal test bed for \cref{finaleq}. Technical details of the simulations and the numerical validation of AdResS w.r.t. the reference fully atomistic simulation are reported in the Appendix.

\subsection{Numerical protocol for the calculation of $\Delta\mu$}

The total chemical potential of a liquid can be separated into the kinetic and potential contributions $\mu = \mu^\id + \mu^\ex$.
In this relation, $\mu^\id$ originates from the probability distribution of the momenta only,
$\propto \exp(-\beta\sum_{i=1}^{N} \vec{p}_{i}^{2}/2m)$, thus it is equivalent to the chemical potential of an ideal gas at the given (uniform) particle density $\rho=N/V$:
\begin{equation}
\mu^\id = \kB T \ln(\rho \Lambda^{3})
\end{equation}
with the thermal wavelength $\Lambda = h/\sqrt{2\pi m\kB T}$. The contribution $\mu^\ex$ is called excess chemical potential and originates from the the position-dependent part of the $N$-particle phase space density $\propto \e^{-\beta U(x_N)}$ (Ref. \citenum{frenkelbook}).
According to the above separation of the chemical potential, \cref{finaleq} is rearranged to:
\begin{equation}
  \Delta\mu= \mu^\ex_{r}-\mu_{0}^\ex+\gamma^\id -\omega_\Delta +\omega_\res
  \label{eq:muatexpanded}
\end{equation}
with $\gamma^\id = \mu_{r}^\id - \mu_{0}^\id = \kB T \log(\rho_{r}/\rho_{0})$, where $\rho_{0}$ is the particle density in the AT region in the initial iteration of the thermodynamic force calculation, that is when no corrections are added to the potential yielding the unbalanced density (\cref{fig:density}).

All the ingredients needed to explicitly calculate $\Delta\mu$, i.e. the unknown perturbation in the chemical potential generated by the thermodynamic force, are now available.
First, $\mu^\ex_{r}$ can be calculated by, e.g., Widom's test particle insertion \cite{tpi} in the fully atomistic simulation of the reference system.
Second, $\mu_{0}^\ex$ instead is the chemical potential the system would have in the AT region if AdResS runs without the thermodynamic force in the transition region $\Delta$.
It can be determined from also Widom's test particle insertion either in the AT region of the initial AdResS set-up or in a standard MD simulation at the density $\rho_0$, that is the density of the AT region in the AdResS set-up without any correction measures.
The latter occurs as the density in the first iteration run for finding the thermodynamic force (see \cref{fig:density} of the Appendix), since we are assuming that the AT region is infinitely large.

\subsection{Numerical results}

All terms contributing to \cref{eq:muatexpanded} can be determined from the fully atomistic simulation of reference ($\mu_r^\ex$ and $\omega_\res$), the AdResS simulation ($\gamma^\id$ and $\omega_\Delta$), and a mix of both simulations ($\mu_0^\ex$) in a straightforward manner as described. Here, after validating the case studies for the AdResS simulation, i.e. investigating their capability for preserving structural and statistical properties of the fluids compared to the reference set-up (see Appendix), we have tested our derivations for four different LJ fluids at different state points (different densities).

\begin{figure}
  \centering\includegraphics[width=\figwidth]{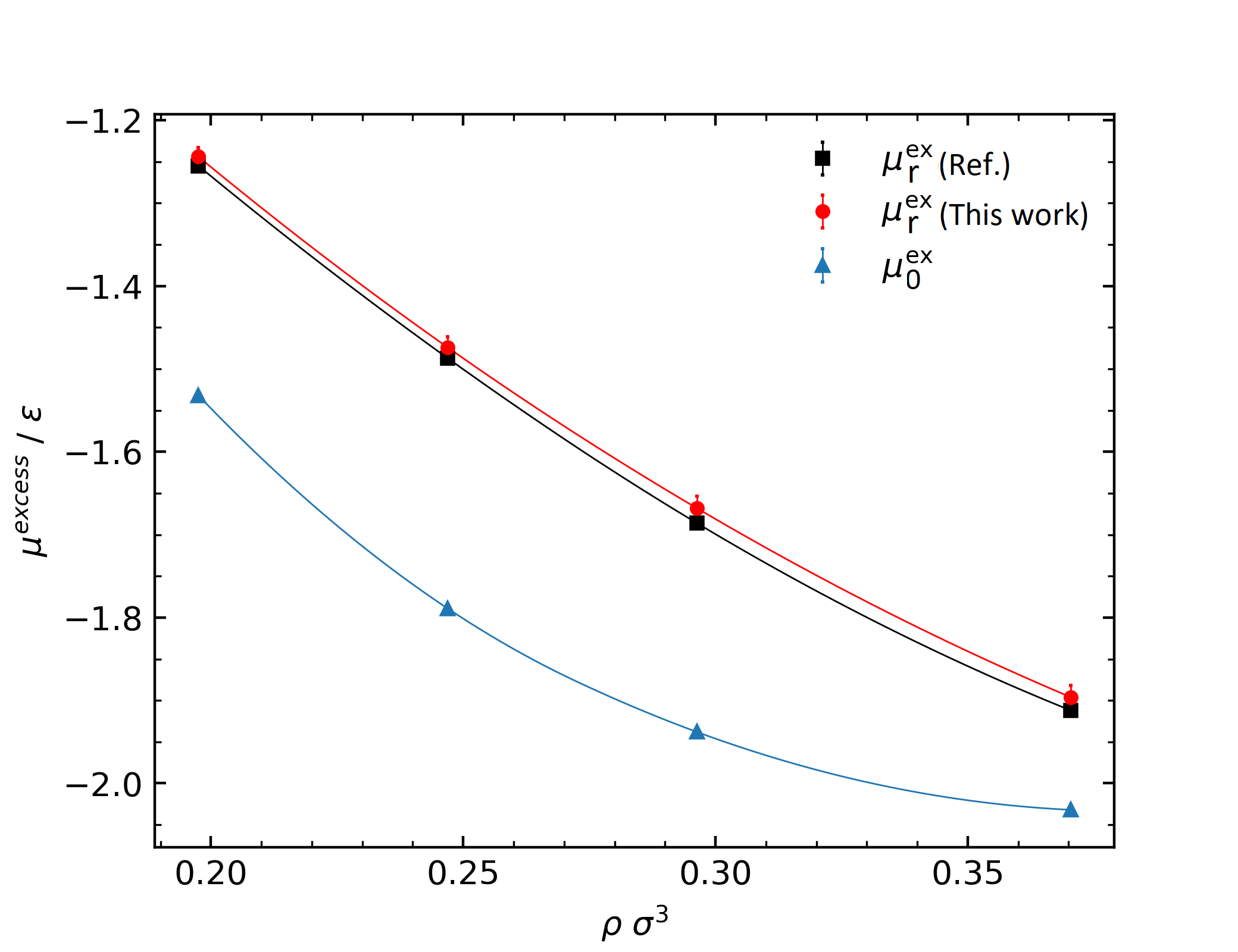}
  \caption{Excess chemical potential of LJ fluids at temperature $T^*=1.5$ as function of the number density $\rho$.
  Values obtained from AdResS simulations (red circles) via \cref{mu_final} are compared to reference data from Widom's test particle insertion in standard MD simulations (black squares).
  The quantity $\mu_0^\ex$ (blue triangles) refers to the AT region of the AdResS set-up with the thermodynamic force  switched off, which results in the modified density $\rho_0$ (see \cref{fig:density}).
  The data points correspond to columns 1, 2, 3, and 8 of \cref{table:mu}.
  }
  \label{fig:mu}
\end{figure}

Simulation results for each contribution to the excess chemical potential relation stated in \cref{eq:muatexpanded} are reported in \cref{table:mu}.
$\Delta \mu$ can be interpreted as the difference between the chemical potential of the fluid within a fully atomistic simulation and the one computed from an AdResS simulation. Interestingly, one would expect that $\Delta\mu\to 0$  as $|\AT|\to\infty$ because the atomistic region would behave as a closed, infinite fully atomistic system with $\mu_{0}\to\mu_{r}$, $\omega_\Delta\to 0$ and $\omega_\res\to 0$. The numerical results of the current simulations are for finite systems with sizes typical of routine AdResS simulations and they actually show that $\Delta\mu\approx 0$ even when $\omega_\Delta$ and $\omega_\res$ are not negligible. This is an interesting result because it allows us to state that the numerical experiments over different densities actually suggest an effective formula:
\begin{equation}
  \mu_{r}-\mu_{0}=\omega_\Delta-\omega_\res \,,
  \label{appeq}
\end{equation}
or,
\begin{equation}
\mu_r^\ex-\mu_0^\ex=-\gamma^\id+\omega_\Delta-\omega_\res \,.
\label{mu_final}
\end{equation}
The relative deviation of $\mu_r^\ex$ calculated from \cref{mu_final} w.r.t. the reference value from fully atomistic simulations, is near or below 1\% in all cases. In particular, the two values coincide within their specified statistical uncertainties (\cref{fig:mu}).
Yet, we note that the reference values systematically lie slightly below the AdResS values, which has a possible source in the neglected contribution due to the capping of unduly large interparticle forces.

We also tested \cref{eq:muatexpanded} for liquid water as a system routinely simulated with AdResS for biological systems such as membranes \cite{physres}. This is a far more complex liquid compared to the Lennard-Jones systems and the simulation set-up is far from mimicking the thermodynamic limit, yet we find that the equation still holds. In this case the dominant correction is $\omega_\res$, while $\Delta\mu$ is comparable with $\gamma^\id$ and $\omega_\Delta$ and these terms contribute with less than  1\% to the sum in \cref{mu_final}.
The possibility to reconstruct the excess chemical potential $\mu_r^\ex$ with high accuracy from an AdResS simulation provides a first-principles confirmation of the physical consistency of AdResS as an open system.
\begin{table*}
% define centred columns of flexible width
\renewcommand\tabularxcolumn[1]{>{\hfill}p{#1}<{\hfill\hbox{}}}
\heavyrulewidth=1pt\lightrulewidth=.5pt\cmidrulewidth\lightrulewidth\cmidrulekern=1ex
% \small
\begin{tabularx}{\textwidth}{cd{5}d{7}d{7}d{7}d{7}@{\extracolsep{.1em}}d{9}d{7}}%@{\extracolsep{.5em}}d{5}}
  \toprule
    \multicolumn{1}{X}{$\rho^*$} &
    \multicolumn{1}{c}{$\mu_r^\ex$} &
    \multicolumn{1}{c}{$\mu_0^\ex$} &
    \multicolumn{1}{c}{$\gamma^\id$} &
    \multicolumn{1}{c}{$\omega_\Delta$} &
    \multicolumn{1}{c}{$\omega_\res$} &
   \multicolumn{1}{c}{$\Delta \mu$} &
    \multicolumn{1}{c}{$\mu_r^\ex$ acc. \cref{mu_final}} \\
  \midrule[\heavyrulewidth]
  0.198 & -1.255(2) & -1.532(3) & -0.385(3) & 0.125(2) & 0.222(3) & -0.011(11) & -1.244(11) \\ % & 0.88\% \\
  0.247 & -1.487(3) & -1.789(4) & -0.411(3) & 0.160(3) & 0.256(3) & -0.013(16) & -1.474(13) \\ % & 0.87\% \\
  0.296 & -1.686(4) & -1.938(5) & -0.384(3) & 0.192(3) & 0.306(3) & -0.018(18) & -1.668(14) \\ % & 1.1\% \\
  0.370 & -1.912(5) & -2.032(5) & -0.268(3) & 0.233(3) & 0.365(3) & -0.016(19) & -1.896(14) \\ % & 0.84\% \\
  \midrule
  \multicolumn{1}{c}{water} &    -24.8(1) &   -21.9(1) & -0.203(3) & 0.210(4) & 3.1(1)  & -0.2(3) & -24.6(3) \\
  \bottomrule
\end{tabularx}
\caption{Breakdown of the chemical potential relation into AdResS-related contributions [\cref{eq:muatexpanded}] for the investigated Lennard-Jones fluids at temperature $T^*=1.5$ and number density as given in the first column.
The values for the density $\rho_0$ (entering $\mu_0^\ex$ and $\gamma^\id$) and the free energy contribution $\omega_\Delta$ related to the thermodynamic force were obtained from AdResS simulations (columns 3 to 5), whereas the results for $\mu_0^\ex$ and $\omega_\res$  (columns 3 and 6) as well as for the reference value for $\mu_r^\ex$ (second column) stem from fully atomistic simulations.
The values for $\mu_r^\ex$ in column~8 were calculated according to \cref{mu_final}.
Chemical potentials and free energies are given in units of $\epsilon$ for the LJ fluids and in units of \si{\kilo\joule\per\mole} for water.
Numbers in parentheses give the uncertainty in the last digit(s).
}
\label{table:mu}
\end{table*}

\section{Conclusions}

We have analyzed the coupling region of the AdResS set-up from the microscopic point of view. We have shown the possibility of explicitly writing the grand potential of the atomistically resolved region in terms of quantities that can be determined from simulations. In particular we have found the relation \eqref{finaleq} between the chemical potential of AdResS and the chemical potential of its reference fully atomistic simulation. The derivation is done under the ideal condition of the thermodynamic limit for the atomistically resolved region, with the coupling conditions considered as small surface effects. The obtained thermodynamic relation was then tested in several numerical experiments, they show that its actual range of validity extends to finite systems with sizes typical of standard AdResS simulations.
Accepting that $\Delta\mu=0$ holds also for a finite (yet not too small) $AT \cup \Delta$ region implies that the equilibrated AdResS (i.e., with $F_\th$ switched on) and the subsystem $S$ of the fully atomistic reference simulation are open systems at different chemical potentials, $\mu_0$ and $\mu_r$, that otherwise exhibit the same physical properties.

The numerical confirmation of the validity of the thermodynamic relations in AdResS provides a statistical mechanics validation of the method as a reasonable numerical approximation of an open system embedded in a reservoir of particles and energy. In conclusion, we have shown that although the abrupt coupling may suggest that a high degree of seemingly artificial conditions are required for the technique to work properly, in effect the numerical conditions are consistent with the statistical mechanics principles of an open system.

\section{Appendix}

\subsection{Technical Details and Validation of $\text{AdResS}$}
For validation of AdResS, a variety of LJ fluids with different state points along with a water model at biological conditions have been studied. The LJ fluid particles are of mass $m$ and interact pairwise with the sharply truncated and shifted LJ potential $U(r) = U_\text{LJ}(r) - U_\text{LJ}(r_c)$ for $r < r_c$ and $U(r) = 0$ otherwise; the cut-off distance was chosen as $r_c = 2.5\sigma$ and the original LJ potential reads
$
 U_\text{LJ}(r) = 4\epsilon \bigl[(r/\sigma)^{-12} - (r/\sigma)^{-6} \bigr] .
$
The parameters $\epsilon$ and $\sigma$ serve as intrinsic units for energy and length, respectively; the unit of time is set to $\tau := \sqrt{m\sigma^2/\epsilon}$. For the case of water, in addition to the mentioned pair interactions, electrostatic potentials are also included with a cut-off radius of \SI{1.2}{nm}.

The LJ fluids were kept at the (dimension-reduced) temperature $T^* := \kB T/\epsilon = 1.5$, which is well above the liquid--vapour critical point, and we investigated four different number densities $\rho^* := \rho \sigma^3 \approx 0.20$, 0.25, 0.30, and 0.37, corresponding to particle numbers $N=\SI{8}{k}$, \SI{10}{k}, \SI{12}{k}, and \SI{15}{k},
where \si{k} stands for the SI prefix for \num{e3}. In the case of water, \num{58990} water molecules (i.e. \num{176970} atoms) at a biological temperature of \SI{323}{K} have been considered for simulations.

In the corresponding AdResS set-ups, the same particle numbers were used for the total of LJ and tracer particles.
The LJ particles were confined to a cuboid simulation box of size $45\sigma\times 30\sigma\times30\sigma$ (for the case of water: $\SI{33.09}{nm}\times \SI{7.37}{nm}\times\SI{7.37}{nm}$), with periodic boundaries imposed at all faces.
For the AdResS set-up, the width of the transition region $\Delta$ along the $x$-axis was set to the cut-off radius, $L_\Delta = r_c$, which provides sufficient space and time for the proper equilibration of particles that entered from the $\Delta$/TR border and changed their resolution abruptly before they reach the AT region of interest.
The width of the AT region was chosen as $L_\text{AT} = 6\sigma$ for LJ cases and $L_\text{AT} = \SI{10}{nm}$ for water simulation, which is small enough to reduce the computational cost significantly compared to a fully atomistic simulation and large enough to be able to mimic and reproduce the thermodynamics and structure of the fluid under study.
The remaining part of the simulation box ($L_\text{TR} = 34\sigma$ for LJ cases and $L_\text{TR} = \SI{20.69}{nm}$ for water) is filled with non-interacting particles (tracers). For the fully atomistic simulations serving as reference, the same geometry of the simulation box was used (\cref{cartoon}a) and observables were computed only in a subvolume of width $L_\text{AT}$ along the $x$-axis, corresponding to the AT region of the AdResS set-up.

All simulations were carried out with the GROMACS software \cite{gromacs} using the stochastic leap-frog integrator with timestep $0.002 \tau$, which acts as a Langevin-type thermostat with the time constant set to $0.05\tau$.
Production runs covered $10^3\tau$ to calculate thermodynamic and statistical properties within the AdResS simulation.
The threshold for capping the force on a particle in the $\Delta$ region was set to $F_\text{cap} = 500\epsilon/\sigma$ and was applied separately for each Cartesian component of the force.
Excess chemical potentials were computed in standard MD simulations using Widom's method \cite{tpi}, in particular, \SI{10}{k} test particles were inserted after each interval of $2\tau$.

\begin{figure*}
  \centering\includegraphics[width=\linewidth]{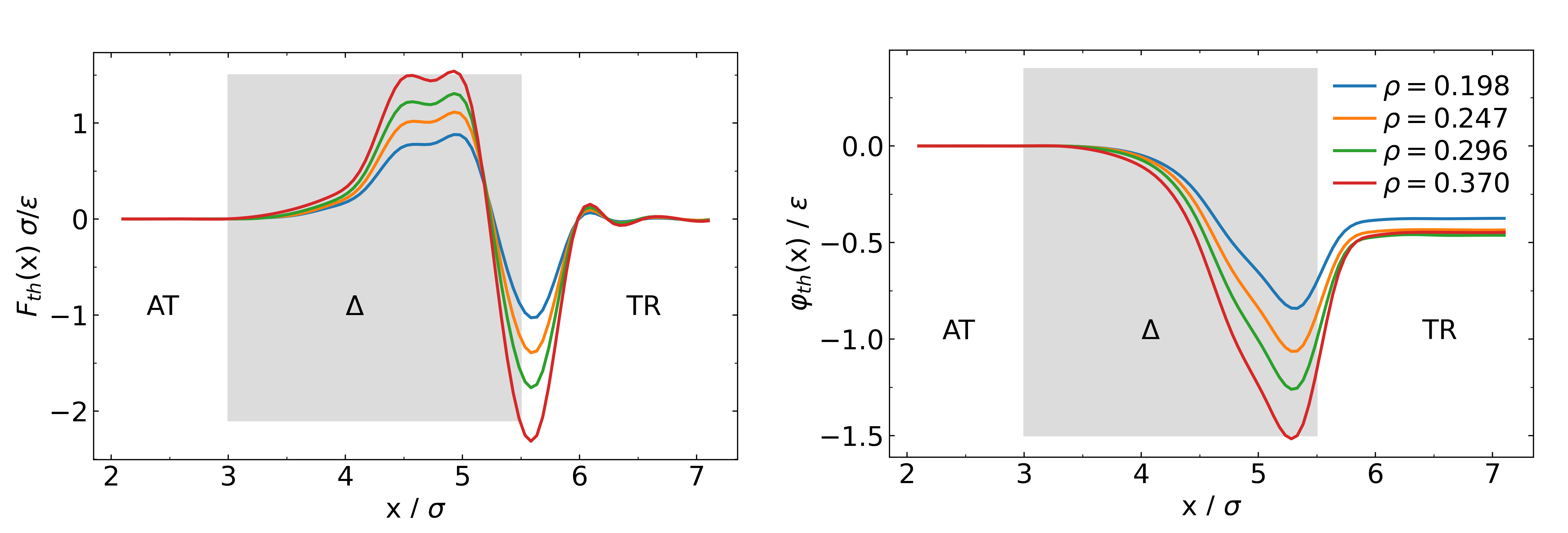}
  \caption{Thermodynamic force $F_\th(x)$ (left) and its potential $\phi_\th(x)$ (right) used in the AdResS set-ups for Lennard-Jones fluids at the same temperature $T^* = 1.5$ and at four reduced densities.
  The thermodynamic force is zero in the AT region by construction and vanishes rapidly inside of the TR region.
  }
  \label{fig:tf}
\end{figure*}

\begin{figure*}
\centering\includegraphics[width=\linewidth]{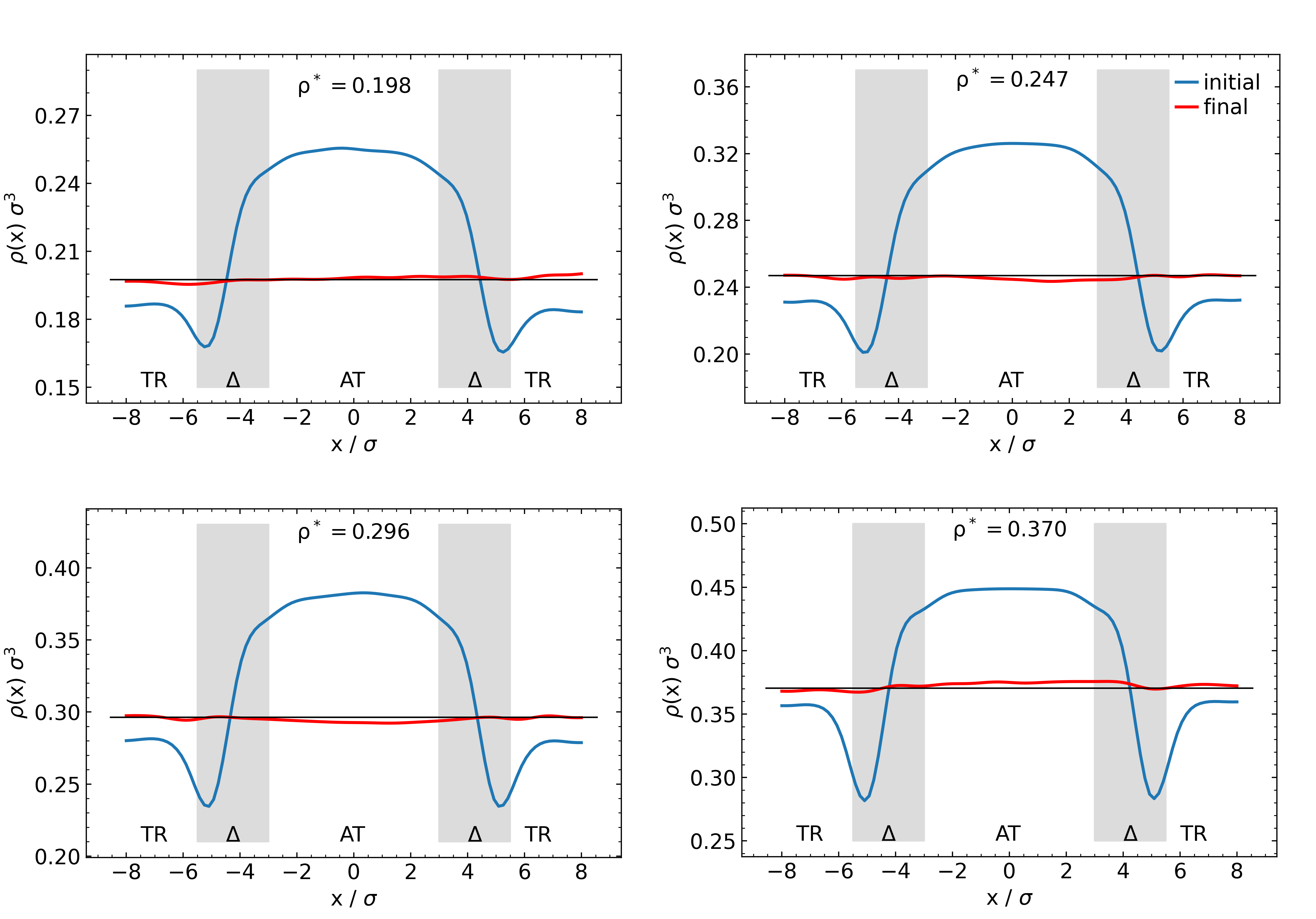}
  \caption{Density profiles $\rho(x)$ across the AdResS set-up along the direction of change of resolution, which is chosen as $x$-axis.
  Lines show the equilibrium profiles generated at the initial and final steps of the iterative calculation of the thermodynamic force $F_\th(x)$.
  The initial choice $F_\th^{(0)}(x)=0$ leads to considerable variations in the density (blue), which are forced to a flat profile (red) within a tolerance of 2\% relative to the constant equilibrium profile $\rho(x) = \rho^*$ (black) by application of the finally obtained $F_\th(x)$ (\cref{fig:tf}).
  The panels show data for Lennard-Jones fluids at the same temperature $T^* = 1.5$ and at four reduced densities as indicated. The transition regions are marked by the gray shadings.
  }
  \label{fig:density}
\end{figure*}

In case of the AdResS set-up and for each density, the thermodynamic force $F_\th(x)$ was calculated iteratively
as described above with the stopping criterion chosen as $\max |\rho(x) - \rho^*| / \rho^* \leq 2\%$; the maximum is taken across the whole simulation box.
The thermodynamic force $F_\th(x)$ was parameterized in terms of a cubic spline interpolation with knot distance $0.3\sigma$.
On average, 10--15 iterations were needed for this scheme to converge, and each iteration involved a simulation run over $200\tau$.

The resulting curves for $F_\th(x)$ are shown in \cref{fig:tf} along with the corresponding potentials $\phi_\th(x)$ obtained from integration of the force.
The main feature of the potentials is a minimum in the $\Delta$ region, close to the $\Delta$/TR boundary ($x=5.5\sigma$), with the depth increasing by a factor of 2 as the density of the fluids is increased from $\rho^*=0.20$ to $\rho^*=0.37$.
Inside the TR region, the potential converges within a distance of $\approx 1\sigma$ from the $\Delta$/TR boundary to a constant $\phi_\text{TR} \approx -0.45\epsilon$, i.e., below the value in the AT region.
The value of $\phi_\text{TR}$ varies only mildly with the density.
Note that its sign is opposite to the case of liquid water at room conditions \cite{advtscomm}.
The physical action of the potential well in $\phi_\th(x)$ is that tracer particles are pulled into the denser fluid in the $\Delta$ region, whereas LJ particles are kept from escaping to the TR region.
Effectively, it yields a flat density profile at the equilibrium density $\rho^*$ of the corresponding LJ fluid, i.e., the AdResS set-up reproduces the density of a fully atomistic reference simulation within the prescribed tolerance (\cref{fig:density}).
In the absence of the thermodynamic force, $F_\th(x) = 0$, the AT and TR regions are unbalanced, generating an excess of particles on one side of the AdResS interface and a depletion on the other.
In the specific examples, the density in the center of the AT region, denoted by $\rho_0$, is increased by 20--30\%, which is compensated by a diminution of the amount of tracer particles.

\begin{figure}
  \centering\includegraphics[width=\figwidth]{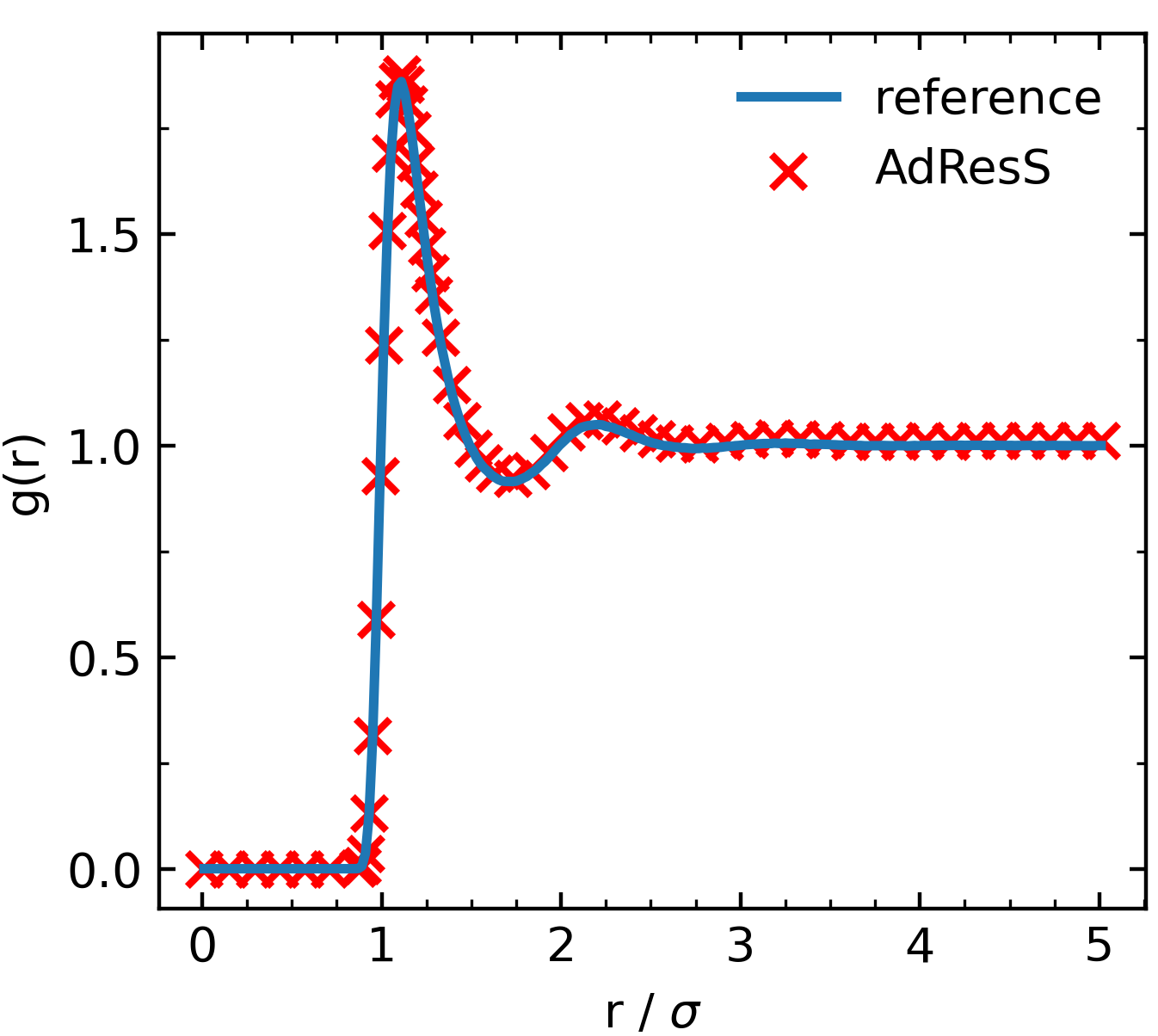}
  \caption{Radial distribution function $g(r)$ obtained from the AT region of the AdResS set-up (red symbols) and the corresponding subvolume of the fully atomistic reference (blue line).
  Data for a Lennard-Jones fluid at temperature $T^* = 1.5$ and number density $\rho^* = 0.37$, using \SI{15}{k} particles in total.
  }
  \label{fig:gr}
\end{figure}

\begin{figure}
  \centering\includegraphics[width=\figwidth]{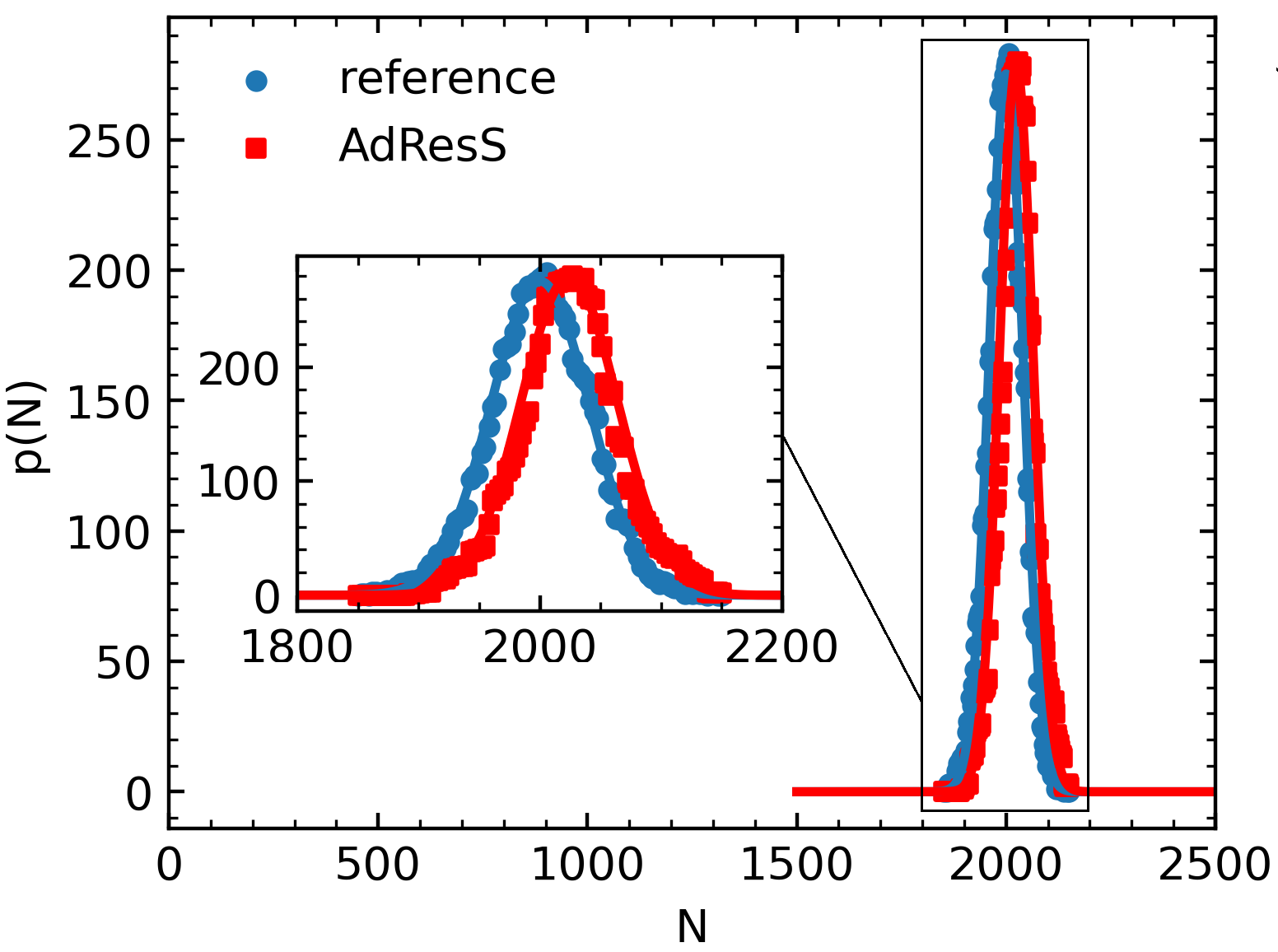}
  \caption{Probability distribution $p(N)$ of finding $N$ particles in the region of interest (AT), which is an open system.
  Comparison of results from the AdResS set-up (red squares) and the fully atomistic reference simulation (blue discs) for a Lennard-Jones fluid at temperature $T^* = 1.5$ and number density $\rho^* = 0.37$.
  Solid lines are fits to a Gaussian distribution.
  The inset shows a close-up of the sharp peak seen in the main panel.
  }
  \label{fig:pn}
\end{figure}

As further checks that the AdResS set-up reproduces the structural and statistical characteristics of the fully atomistic simulation, we compared the radial distribution function $g(r)$ from both approaches, which yield a perfect match (data for $\rho^*=0.37$ are shown in \cref{fig:gr}).
Second, we tested the permeability of the AT/$\Delta$ boundary by inspecting the probability distribution $p(N)$ for finding $N$ particles in the AT region and in the corresponding subvolume of the fully atomistic simulation (\cref{fig:pn}).
Both distributions superpose closely and resemble a Gaussian; the small shift of the peak positions is related to the allowed tolerance on $\rho(x)$ in the computation of the thermodynamic force. For the density $\rho_\at^* = 0.37$, we obtained mean values $\expect{N} = \num{2000}$ and $\num{2024}$ for the reference and for AdResS, respectively. Similarly, the standard deviations $\std(N) = 40.6$ and $40.1$, being a measure of the compressibility, differ by only 1.3\%.
We conclude that the AT region of the AdResS set-up used here is a good representation of an open subvolume of a fully atomistic simulation.

\subsection{{The capped energy} is negligible}
\begin{figure}
  \centering\includegraphics[width=\figwidth]{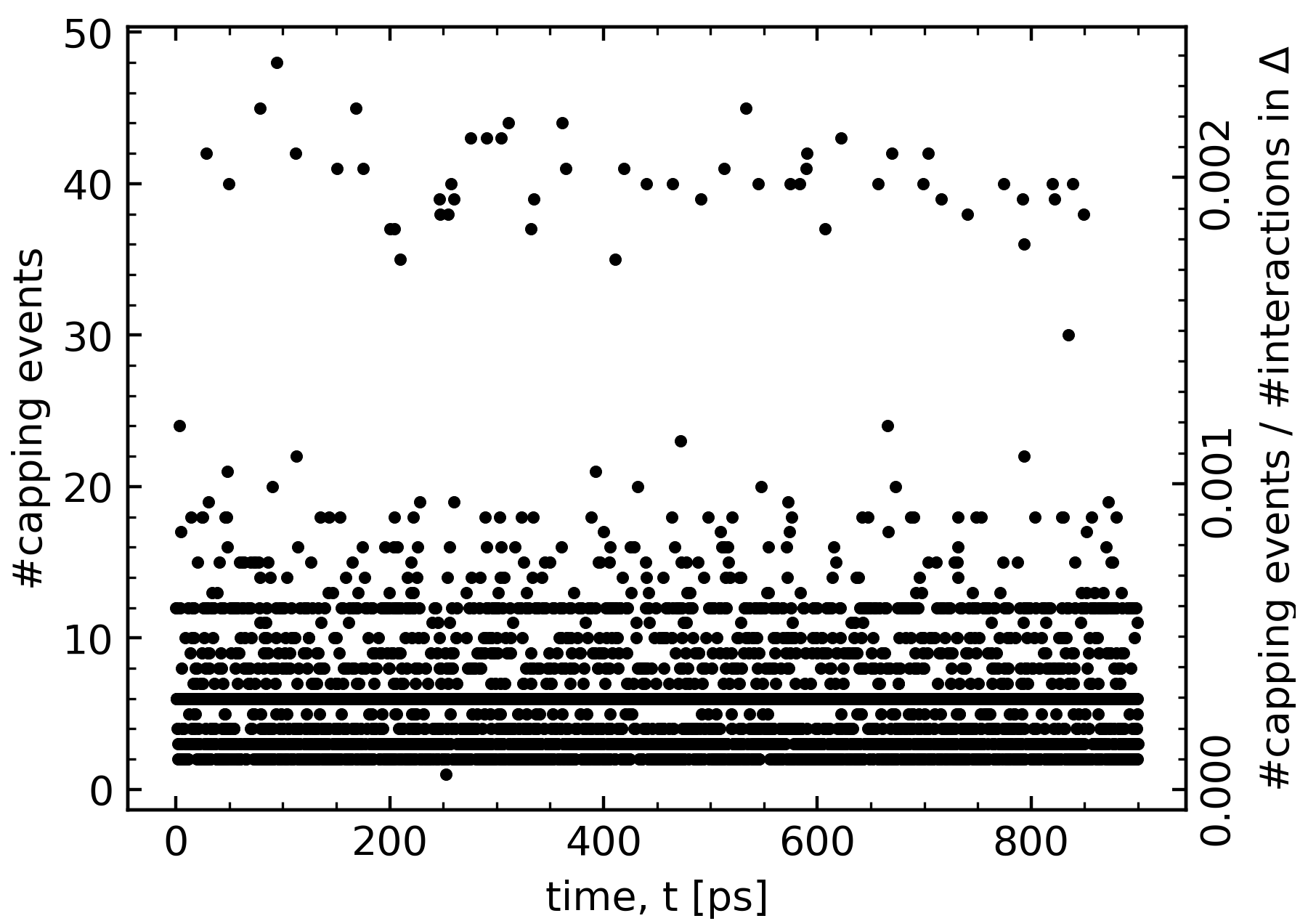}
  \caption{Number of incidences of force capping per MD integration step, relative to the total number of pair interactions in the transition region $\Delta$ as a function of time,
  the latter number was estimated to \num{2e4} for the LJ fluid at the density $\rho^*=0.37$.}
  \label{fig:capping}
\end{figure}

The {force capping} acting in the $\Delta$ region takes care of the divergent interaction potentials, which is technically needed due to the sudden introduction of new interactions upon tracer particles entering the atomistically resolved region. Given a certain configuration of molecules in the $\AT\cup\Delta$ region, the force capping would renormalize the interaction energy of two molecules, located at the very interface between the $\Delta$ region and the tracer region, which have a distance that cannot occur in a fully atomistic simulation.
However, this term is negligible compared to the other contributions as evidenced numerically for the LJ fluid at the density $\rho^*=0.37$, which exhibits the highest frequency of force capping incidences in this study (\cref{fig:capping}).
The number of incidences of force cappings rarely exceeds a value of 20 in each MD integration step, which is three orders of magnitude smaller than the total number of pair interactions in the $\Delta$ region, estimated to \num{2e4} based on the particle density and the radial distribution function (\cref{fig:gr}).
Furthermore, the capping is equivalent to a global modification of the highly repulsive part of the interaction potentials, which has marginal repercussions on the physical properties of the fluid;
specifically for the LJ potential and the choice for $F_\text{cap}=500 \sigma / \epsilon$ used here, the capping corresponds to a modification of the potential for distances shorter than $r_\text{cap} \approx 0.82\,\sigma$ or potential energies $U(r) \gtrsim 28\, \epsilon$.

\section*{Acknowledgments} 
This research has been funded by Deutsche Forschungsgemeinschaft (DFG) through grant CRC 1114 ``Scaling Cascade in Complex Systems,'' Project Number 235221301, Project C01 ``Adaptive coupling of scales in molecular dynamics and beyond to fluid dynamics.''
We thank John Whittaker for helping to set up the numerical simulations.

\bibliography{paper}

%Control: production of eprint (-1) disabled
\begin{thebibliography}{23}%
\makeatletter
\providecommand \@ifxundefined [1]{%
 \@ifx{#1\undefined}
}%
\providecommand \@ifnum [1]{%
 \ifnum #1\expandafter \@firstoftwo
 \else \expandafter \@secondoftwo
 \fi
}%
\providecommand \@ifx [1]{%
 \ifx #1\expandafter \@firstoftwo
 \else \expandafter \@secondoftwo
 \fi
}%
\providecommand \natexlab [1]{#1}%
\providecommand \enquote  [1]{``#1''}%
\providecommand \bibnamefont  [1]{#1}%
\providecommand \bibfnamefont [1]{#1}%
\providecommand \citenamefont [1]{#1}%
\providecommand \href@noop [0]{\@secondoftwo}%
\providecommand \href [0]{\begingroup \@sanitize@url \@href}%
\providecommand \@href[1]{\@@startlink{#1}\@@href}%
\providecommand \@@href[1]{\endgroup#1\@@endlink}%
\providecommand \@sanitize@url [0]{\catcode `\\12\catcode `\$12\catcode
  `\&12\catcode `\#12\catcode `\^12\catcode `\_12\catcode `\%12\relax}%
\providecommand \@@startlink[1]{}%
\providecommand \@@endlink[0]{}%
\providecommand \url  [0]{\begingroup\@sanitize@url \@url }%
\providecommand \@url [1]{\endgroup\@href {#1}{\urlprefix }}%
\providecommand \urlprefix  [0]{URL }%
\providecommand \Eprint [0]{\href }%
\providecommand \doibase [0]{http://dx.doi.org/}%
\providecommand \selectlanguage [0]{\@gobble}%
\providecommand \bibinfo  [0]{\@secondoftwo}%
\providecommand \bibfield  [0]{\@secondoftwo}%
\providecommand \translation [1]{[#1]}%
\providecommand \BibitemOpen [0]{}%
\providecommand \bibitemStop [0]{}%
\providecommand \bibitemNoStop [0]{.\EOS\space}%
\providecommand \EOS [0]{\spacefactor3000\relax}%
\providecommand \BibitemShut  [1]{\csname bibitem#1\endcsname}%
\let\auto@bib@innerbib\@empty
%</preamble>
\bibitem [{\citenamefont {Praprotnik}, \citenamefont {Delle~Site},\ and\
  \citenamefont {Kremer}(2005)}]{jcpor}%
  \BibitemOpen
  \bibfield  {author} {\bibinfo {author} {\bibfnamefont {M.}~\bibnamefont
  {Praprotnik}}, \bibinfo {author} {\bibfnamefont {L.}~\bibnamefont
  {Delle~Site}}, \ and\ \bibinfo {author} {\bibfnamefont {K.}~\bibnamefont
  {Kremer}},\ }\bibfield  {title} {\enquote {\bibinfo {title} {Adaptive
  resolution molecular-dynamics simulation: Changing the degrees of freedom on
  the fly},}\ }\href@noop {} {\bibfield  {journal} {\bibinfo  {journal} {J.
  Chem. Phys.}\ }\textbf {\bibinfo {volume} {123}},\ \bibinfo {pages} {224106}
  (\bibinfo {year} {2005})}\BibitemShut {NoStop}%
\bibitem [{\citenamefont {Praprotnik}, \citenamefont {Delle~Site},\ and\
  \citenamefont {Kremer}(2008)}]{annurev}%
  \BibitemOpen
  \bibfield  {author} {\bibinfo {author} {\bibfnamefont {M.}~\bibnamefont
  {Praprotnik}}, \bibinfo {author} {\bibfnamefont {L.}~\bibnamefont
  {Delle~Site}}, \ and\ \bibinfo {author} {\bibfnamefont {K.}~\bibnamefont
  {Kremer}},\ }\bibfield  {title} {\enquote {\bibinfo {title} {Multiscale
  simulation of soft matter: From scale bridging to adaptive resolution},}\
  }\href@noop {} {\bibfield  {journal} {\bibinfo  {journal} {Annu. Rev. Phys.
  Chem.}\ }\textbf {\bibinfo {volume} {59}},\ \bibinfo {pages} {545--571}
  (\bibinfo {year} {2008})}\BibitemShut {NoStop}%
\bibitem [{\citenamefont {Delle~Site}\ and\ \citenamefont
  {Praprotnik}(2017)}]{physrep}%
  \BibitemOpen
  \bibfield  {author} {\bibinfo {author} {\bibfnamefont {L.}~\bibnamefont
  {Delle~Site}}\ and\ \bibinfo {author} {\bibfnamefont {M.}~\bibnamefont
  {Praprotnik}},\ }\bibfield  {title} {\enquote {\bibinfo {title} {Molecular
  systems with open boundaries: Theory and simulation},}\ }\href@noop {}
  {\bibfield  {journal} {\bibinfo  {journal} {Phys.Rep.}\ }\textbf {\bibinfo
  {volume} {693}},\ \bibinfo {pages} {1--56} (\bibinfo {year}
  {2017})}\BibitemShut {NoStop}%
\bibitem [{\citenamefont {Lebowitz}\ and\ \citenamefont
  {Bergmann}(1957)}]{leb1}%
  \BibitemOpen
  \bibfield  {author} {\bibinfo {author} {\bibfnamefont {J.}~\bibnamefont
  {Lebowitz}}\ and\ \bibinfo {author} {\bibfnamefont {P.}~\bibnamefont
  {Bergmann}},\ }\bibfield  {title} {\enquote {\bibinfo {title} {Irreversible
  {G}ibbsian ensembles},}\ }\href@noop {} {\bibfield  {journal} {\bibinfo
  {journal} {Ann. Phys.}\ }\textbf {\bibinfo {volume} {1}},\ \bibinfo {pages}
  {1} (\bibinfo {year} {1957})}\BibitemShut {NoStop}%
\bibitem [{\citenamefont {Bergmann}\ and\ \citenamefont
  {Lebowitz}(1955)}]{leb2}%
  \BibitemOpen
  \bibfield  {author} {\bibinfo {author} {\bibfnamefont {P.}~\bibnamefont
  {Bergmann}}\ and\ \bibinfo {author} {\bibfnamefont {J.}~\bibnamefont
  {Lebowitz}},\ }\bibfield  {title} {\enquote {\bibinfo {title} {New approach
  to nonequilibrium processes},}\ }\href@noop {} {\bibfield  {journal}
  {\bibinfo  {journal} {Phys. Rev.}\ }\textbf {\bibinfo {volume} {99}},\
  \bibinfo {pages} {578} (\bibinfo {year} {1955})}\BibitemShut {NoStop}%
\bibitem [{\citenamefont {Agarwal}\ \emph {et~al.}(2015)\citenamefont
  {Agarwal}, \citenamefont {Zhu}, \citenamefont {Hartmann}, \citenamefont
  {Wang},\ and\ \citenamefont {Delle~Site}}]{njp}%
  \BibitemOpen
  \bibfield  {author} {\bibinfo {author} {\bibfnamefont {A.}~\bibnamefont
  {Agarwal}}, \bibinfo {author} {\bibfnamefont {J.}~\bibnamefont {Zhu}},
  \bibinfo {author} {\bibfnamefont {C.}~\bibnamefont {Hartmann}}, \bibinfo
  {author} {\bibfnamefont {H.}~\bibnamefont {Wang}}, \ and\ \bibinfo {author}
  {\bibfnamefont {L.}~\bibnamefont {Delle~Site}},\ }\bibfield  {title}
  {\enquote {\bibinfo {title} {Molecular dynamics in a grand ensemble:
  {B}ergmann-{L}ebowitz model and adaptive resolution simulation},}\
  }\href@noop {} {\bibfield  {journal} {\bibinfo  {journal} {New. J. Phys.}\
  }\textbf {\bibinfo {volume} {17}},\ \bibinfo {pages} {083042} (\bibinfo
  {year} {2015})}\BibitemShut {NoStop}%
\bibitem [{\citenamefont {Delle~Site}(2016)}]{preliouv}%
  \BibitemOpen
  \bibfield  {author} {\bibinfo {author} {\bibfnamefont {L.}~\bibnamefont
  {Delle~Site}},\ }\bibfield  {title} {\enquote {\bibinfo {title} {Formulation
  of liouville's theorem for grand ensemble molecular simulations},}\
  }\href@noop {} {\bibfield  {journal} {\bibinfo  {journal} {Phys.Rev.E}\
  }\textbf {\bibinfo {volume} {93}},\ \bibinfo {pages} {022130} (\bibinfo
  {year} {2016})}\BibitemShut {NoStop}%
\bibitem [{\citenamefont {Krekeler}\ \emph {et~al.}(2018)\citenamefont
  {Krekeler}, \citenamefont {Agarwal}, \citenamefont {Junghans}, \citenamefont
  {Praprotnik}, ,\ and\ \citenamefont {Delle~Site}}]{krekjcp}%
  \BibitemOpen
  \bibfield  {author} {\bibinfo {author} {\bibfnamefont {C.}~\bibnamefont
  {Krekeler}}, \bibinfo {author} {\bibfnamefont {A.}~\bibnamefont {Agarwal}},
  \bibinfo {author} {\bibfnamefont {C.}~\bibnamefont {Junghans}}, \bibinfo
  {author} {\bibfnamefont {M.}~\bibnamefont {Praprotnik}}, , \ and\ \bibinfo
  {author} {\bibfnamefont {L.}~\bibnamefont {Delle~Site}},\ }\bibfield  {title}
  {\enquote {\bibinfo {title} {Adaptive resolution molecular dynamics
  technique: Down to the essential},}\ }\href@noop {} {\bibfield  {journal}
  {\bibinfo  {journal} {J.Chem.Phys.}\ }\textbf {\bibinfo {volume} {149}},\
  \bibinfo {pages} {24104} (\bibinfo {year} {2018})}\BibitemShut {NoStop}%
\bibitem [{\citenamefont {Delle~Site}\ \emph {et~al.}(2019)\citenamefont
  {Delle~Site}, \citenamefont {Krekeler}, \citenamefont {Whittaker},
  \citenamefont {Agarwal}, \citenamefont {Klein},\ and\ \citenamefont
  {H\"{o}fling}}]{advtscomm}%
  \BibitemOpen
  \bibfield  {author} {\bibinfo {author} {\bibfnamefont {L.}~\bibnamefont
  {Delle~Site}}, \bibinfo {author} {\bibfnamefont {C.}~\bibnamefont
  {Krekeler}}, \bibinfo {author} {\bibfnamefont {J.}~\bibnamefont {Whittaker}},
  \bibinfo {author} {\bibfnamefont {A.}~\bibnamefont {Agarwal}}, \bibinfo
  {author} {\bibfnamefont {R.}~\bibnamefont {Klein}}, \ and\ \bibinfo {author}
  {\bibfnamefont {F.}~\bibnamefont {H\"{o}fling}},\ }\bibfield  {title}
  {\enquote {\bibinfo {title} {Molecular dynamics of open systems: construction
  of a mean-field particle reservoir},}\ }\href@noop {} {\bibfield  {journal}
  {\bibinfo  {journal} {Adv. Theory Simul.}\ }\textbf {\bibinfo {volume} {2}},\
  \bibinfo {pages} {1900014} (\bibinfo {year} {2019})}\BibitemShut {NoStop}%
\bibitem [{\citenamefont {Delle~Site}\ and\ \citenamefont {Klein}(2020)}]{jmp}%
  \BibitemOpen
  \bibfield  {author} {\bibinfo {author} {\bibfnamefont {L.}~\bibnamefont
  {Delle~Site}}\ and\ \bibinfo {author} {\bibfnamefont {R.}~\bibnamefont
  {Klein}},\ }\bibfield  {title} {\enquote {\bibinfo {title} {Liouville-type
  equation for the $n$-particle distribution function of an open system},}\
  }\href@noop {} {\bibfield  {journal} {\bibinfo  {journal} {J.Math.Phys.}\
  }\textbf {\bibinfo {volume} {61}},\ \bibinfo {pages} {083102} (\bibinfo
  {year} {2020})}\BibitemShut {NoStop}%
\bibitem [{\citenamefont {Ebrahimi~Viand}\ \emph {et~al.}(2020)\citenamefont
  {Ebrahimi~Viand}, \citenamefont {H\"{o}fling}, \citenamefont {Klein},\ and\
  \citenamefont {Delle~Site}}]{noneq1}%
  \BibitemOpen
  \bibfield  {author} {\bibinfo {author} {\bibfnamefont {R.}~\bibnamefont
  {Ebrahimi~Viand}}, \bibinfo {author} {\bibfnamefont {F.}~\bibnamefont
  {H\"{o}fling}}, \bibinfo {author} {\bibfnamefont {R.}~\bibnamefont {Klein}},
  \ and\ \bibinfo {author} {\bibfnamefont {L.}~\bibnamefont {Delle~Site}},\
  }\bibfield  {title} {\enquote {\bibinfo {title} {Theory and simulation of
  open systems out of equilibrium},}\ }\href@noop {} {\bibfield  {journal}
  {\bibinfo  {journal} {J.Chem.Phys.}\ }\textbf {\bibinfo {volume} {153}},\
  \bibinfo {pages} {101102} (\bibinfo {year} {2020})}\BibitemShut {NoStop}%
\bibitem [{\citenamefont {Espa\~nol}\ \emph {et~al.}(2015)\citenamefont
  {Espa\~nol}, \citenamefont {Delgado-Buscalioni}, \citenamefont {Everaers},
  \citenamefont {Potestio}, \citenamefont {Donadio},\ and\ \citenamefont
  {Kremer}}]{prlraff}%
  \BibitemOpen
  \bibfield  {author} {\bibinfo {author} {\bibfnamefont {P.}~\bibnamefont
  {Espa\~nol}}, \bibinfo {author} {\bibfnamefont {R.}~\bibnamefont
  {Delgado-Buscalioni}}, \bibinfo {author} {\bibfnamefont {R.}~\bibnamefont
  {Everaers}}, \bibinfo {author} {\bibfnamefont {R.}~\bibnamefont {Potestio}},
  \bibinfo {author} {\bibfnamefont {D.}~\bibnamefont {Donadio}}, \ and\
  \bibinfo {author} {\bibfnamefont {K.}~\bibnamefont {Kremer}},\ }\bibfield
  {title} {\enquote {\bibinfo {title} {Statistical mechanics of hamiltonian
  adaptive resolution simulations},}\ }\href@noop {} {\bibfield  {journal}
  {\bibinfo  {journal} {J. Chem. Phys.}\ }\textbf {\bibinfo {volume} {142}},\
  \bibinfo {pages} {064115} (\bibinfo {year} {2015})}\BibitemShut {NoStop}%
\bibitem [{\citenamefont {Praprotnik}\ \emph {et~al.}(2011)\citenamefont
  {Praprotnik}, \citenamefont {Poblete}, \citenamefont {Delle~Site},\ and\
  \citenamefont {Kremer}}]{prlmatej}%
  \BibitemOpen
  \bibfield  {author} {\bibinfo {author} {\bibfnamefont {M.}~\bibnamefont
  {Praprotnik}}, \bibinfo {author} {\bibfnamefont {S.}~\bibnamefont {Poblete}},
  \bibinfo {author} {\bibfnamefont {L.}~\bibnamefont {Delle~Site}}, \ and\
  \bibinfo {author} {\bibfnamefont {K.}~\bibnamefont {Kremer}},\ }\bibfield
  {title} {\enquote {\bibinfo {title} {Comment on "{A}daptive multiscale
  molecular dynamics of macromolecular fluids"},}\ }\href@noop {} {\bibfield
  {journal} {\bibinfo  {journal} {Phys. Rev. Lett.}\ }\textbf {\bibinfo
  {volume} {107}},\ \bibinfo {pages} {099801} (\bibinfo {year}
  {2011})}\BibitemShut {NoStop}%
\bibitem [{\citenamefont {Wang}, \citenamefont {Sch\"{u}tte},\ and\
  \citenamefont {Delle~Site}(2012)}]{jctchan}%
  \BibitemOpen
  \bibfield  {author} {\bibinfo {author} {\bibfnamefont {H.}~\bibnamefont
  {Wang}}, \bibinfo {author} {\bibfnamefont {C.}~\bibnamefont {Sch\"{u}tte}}, \
  and\ \bibinfo {author} {\bibfnamefont {L.}~\bibnamefont {Delle~Site}},\
  }\bibfield  {title} {\enquote {\bibinfo {title} {Adaptive resolution
  simulation (adress): A smooth thermodynamic and structural transition from
  atomistic to coarse grained resolution and vice versa in a grand canonical
  fashion},}\ }\href@noop {} {\bibfield  {journal} {\bibinfo  {journal} {J.
  Chem. Th.Comp.}\ }\textbf {\bibinfo {volume} {8}},\ \bibinfo {pages} {2878}
  (\bibinfo {year} {2012})}\BibitemShut {NoStop}%
\bibitem [{\citenamefont {Wang}\ \emph {et~al.}(2013)\citenamefont {Wang},
  \citenamefont {Hartmann}, \citenamefont {Sch\"{u}tte},\ and\ \citenamefont
  {Delle~Site}}]{prx}%
  \BibitemOpen
  \bibfield  {author} {\bibinfo {author} {\bibfnamefont {H.}~\bibnamefont
  {Wang}}, \bibinfo {author} {\bibfnamefont {C.}~\bibnamefont {Hartmann}},
  \bibinfo {author} {\bibfnamefont {C.}~\bibnamefont {Sch\"{u}tte}}, \ and\
  \bibinfo {author} {\bibfnamefont {L.}~\bibnamefont {Delle~Site}},\ }\bibfield
   {title} {\enquote {\bibinfo {title} {Grand-canonical-like molecular-dynamics
  simulations by using an adaptive-resolution technique},}\ }\href@noop {}
  {\bibfield  {journal} {\bibinfo  {journal} {Phys. Rev. X}\ }\textbf {\bibinfo
  {volume} {3}},\ \bibinfo {pages} {011018} (\bibinfo {year}
  {2013})}\BibitemShut {NoStop}%
\bibitem [{\citenamefont {Fritsch}\ \emph {et~al.}(2012)\citenamefont
  {Fritsch}, \citenamefont {Poblete}, \citenamefont {Junghans}, \citenamefont
  {Ciccotti}, \citenamefont {Delle~Site},\ and\ \citenamefont
  {Kremer}}]{prl2012}%
  \BibitemOpen
  \bibfield  {author} {\bibinfo {author} {\bibfnamefont {S.}~\bibnamefont
  {Fritsch}}, \bibinfo {author} {\bibfnamefont {S.}~\bibnamefont {Poblete}},
  \bibinfo {author} {\bibfnamefont {C.}~\bibnamefont {Junghans}}, \bibinfo
  {author} {\bibfnamefont {G.}~\bibnamefont {Ciccotti}}, \bibinfo {author}
  {\bibfnamefont {L.}~\bibnamefont {Delle~Site}}, \ and\ \bibinfo {author}
  {\bibfnamefont {K.}~\bibnamefont {Kremer}},\ }\bibfield  {title} {\enquote
  {\bibinfo {title} {Adaptive resolution molecular dynamics simulation through
  coupling to an internal particle reservoir},}\ }\href@noop {} {\bibfield
  {journal} {\bibinfo  {journal} {Phys. Rev. Lett.}\ }\textbf {\bibinfo
  {volume} {108}},\ \bibinfo {pages} {170602} (\bibinfo {year}
  {2012})}\BibitemShut {NoStop}%
\bibitem [{\citenamefont {Ciccotti}\ and\ \citenamefont
  {Delle~Site}(2019)}]{softmatt}%
  \BibitemOpen
  \bibfield  {author} {\bibinfo {author} {\bibfnamefont {G.}~\bibnamefont
  {Ciccotti}}\ and\ \bibinfo {author} {\bibfnamefont {L.}~\bibnamefont
  {Delle~Site}},\ }\bibfield  {title} {\enquote {\bibinfo {title} {The physics
  of open systems for the simulation of complex molecular environments in soft
  matter},}\ }\href@noop {} {\bibfield  {journal} {\bibinfo  {journal} {Soft
  Matter}\ }\textbf {\bibinfo {volume} {15}},\ \bibinfo {pages} {2114}
  (\bibinfo {year} {2019})}\BibitemShut {NoStop}%
\bibitem [{\citenamefont {Höf{}ling}\ and\ \citenamefont
  {Dietrich}(2020)}]{jcp-fscorr}%
  \BibitemOpen
  \bibfield  {author} {\bibinfo {author} {\bibfnamefont {F.}~\bibnamefont
  {Höf{}ling}}\ and\ \bibinfo {author} {\bibfnamefont {S.}~\bibnamefont
  {Dietrich}},\ }\bibfield  {title} {\enquote {\bibinfo {title} {Finite-size
  corrections for the static structure factor of a liquid slab with open
  boundaries},}\ }\href@noop {} {\bibfield  {journal} {\bibinfo  {journal} {J.
  Chem. Phys.}\ }\textbf {\bibinfo {volume} {153}},\ \bibinfo {pages} {054119}
  (\bibinfo {year} {2020})}\BibitemShut {NoStop}%
\bibitem [{\citenamefont {Huang}(1986)}]{huang}%
  \BibitemOpen
  \bibfield  {author} {\bibinfo {author} {\bibfnamefont {K.}~\bibnamefont
  {Huang}},\ }\href@noop {} {\emph {\bibinfo {title} {Statistical mechanics}}}\
  (\bibinfo  {publisher} {Wiley},\ \bibinfo {year} {1986})\BibitemShut
  {NoStop}%
\bibitem [{\citenamefont {Frenkel}\ and\ \citenamefont
  {Smit}(1996)}]{frenkelbook}%
  \BibitemOpen
  \bibfield  {author} {\bibinfo {author} {\bibfnamefont {D.}~\bibnamefont
  {Frenkel}}\ and\ \bibinfo {author} {\bibfnamefont {B.}~\bibnamefont {Smit}},\
  }\href@noop {} {\emph {\bibinfo {title} {Understanding Molecular
  Simulation}}}\ (\bibinfo  {publisher} {Academic Press},\ \bibinfo {year}
  {1996})\BibitemShut {NoStop}%
\bibitem [{\citenamefont {Widom}(1963)}]{tpi}%
  \BibitemOpen
  \bibfield  {author} {\bibinfo {author} {\bibfnamefont {B.}~\bibnamefont
  {Widom}},\ }\bibfield  {title} {\enquote {\bibinfo {title} {Some topics in
  the theory of fluids},}\ }\href@noop {} {\bibfield  {journal} {\bibinfo
  {journal} {J.Chem.Phys.}\ }\textbf {\bibinfo {volume} {39}},\ \bibinfo
  {pages} {2808--2812} (\bibinfo {year} {1963})}\BibitemShut {NoStop}%
\bibitem [{\citenamefont {Whittaker}\ and\ \citenamefont
  {Delle~Site}(2019)}]{physres}%
  \BibitemOpen
  \bibfield  {author} {\bibinfo {author} {\bibfnamefont {J.}~\bibnamefont
  {Whittaker}}\ and\ \bibinfo {author} {\bibfnamefont {L.}~\bibnamefont
  {Delle~Site}},\ }\bibfield  {title} {\enquote {\bibinfo {title}
  {Investigation of the hydration shell of a membrane in an open system
  molecular dynamics simulation},}\ }\href@noop {} {\bibfield  {journal}
  {\bibinfo  {journal} {Phys.Rev.Res.}\ }\textbf {\bibinfo {volume} {1}},\
  \bibinfo {pages} {033099} (\bibinfo {year} {2019})}\BibitemShut {NoStop}%
\bibitem [{\citenamefont {M.J.Abraham}\ \emph {et~al.}(2015)\citenamefont
  {M.J.Abraham}, \citenamefont {T.Murtola}, \citenamefont {R.Schulz},
  \citenamefont {S.Pall}, \citenamefont {Smith}, \citenamefont {B.Hess},\ and\
  \citenamefont {E.Lindahl}}]{gromacs}%
  \BibitemOpen
  \bibfield  {author} {\bibinfo {author} {\bibnamefont {M.J.Abraham}}, \bibinfo
  {author} {\bibnamefont {T.Murtola}}, \bibinfo {author} {\bibnamefont
  {R.Schulz}}, \bibinfo {author} {\bibnamefont {S.Pall}}, \bibinfo {author}
  {\bibfnamefont {J.}~\bibnamefont {Smith}}, \bibinfo {author} {\bibnamefont
  {B.Hess}}, \ and\ \bibinfo {author} {\bibnamefont {E.Lindahl}},\ }\bibfield
  {title} {\enquote {\bibinfo {title} {Gromacs: High performance molecular
  simulations through multi-level parallelism from laptops to
  supercomputers},}\ }\href@noop {} {\bibfield  {journal} {\bibinfo  {journal}
  {SoftwareX}\ }\textbf {\bibinfo {volume} {1-2}},\ \bibinfo {pages} {19 -- 25}
  (\bibinfo {year} {2015})}\BibitemShut {NoStop}%
\end{thebibliography}%
\end{document}